\documentclass[a4paper, 10pt, twocolumn]{article}
\usepackage{epsfig}

\begin{document}

\begin{titlepage}

\begin{flushright}{Edinburgh 2000-26 \\ Dec 2000}
\end{flushright}

\vspace{2.6cm}
\begin{center}
{\Large \bf Multianode Photo Multipliers for Ring Imaging\\ 
Cherenkov Detectors}

\vspace*{1.0cm}

{\Large \bf Franz Muheim}

\vspace*{1.0cm}

{\it The University of Edinburgh, 
        Department of Physics and Astronomy\\
        Mayfield Road, Edinburgh EH9 3JZ, Scotland/UK\\
        E-mail: F.Muheim@ed.ac.uk}
\end{center}

\vspace*{1.0cm}

\begin{abstract}
The  64-channel Multianode Photo Multiplier has been evaluated 
as a possible choice for the photo detectors of the LHCb Ring Imaging Cherenkov
detector.
\end{abstract}

\vfill
\begin{center}
{\it Contributed to the Proceedings of the 30$^{th}$ International 
Conference on High Energy Physics, \\
7/27/2000---8/2/2000, Osaka, Japan}

\end{center}

\end{titlepage}

\section{Introduction}
The LHCb experiment  will exploit the large rates of 
$B$ hadrons that will be produced at the Large Hadron Collider
and make precision measurements of CP violation.
Excellent particle identification is needed for 
LHCb, e.g. three kaons in a large momentum range are produced by 
the decay $B^0_s \to D_s^{\mp} K^{\pm}$, 
$D_s^+ \to \phi \pi^+$, $\phi \to K^+ K^-$ 
which is sensitive to the CP violating phase $\gamma$.
Charged particles will be identified by means of two
Ring Imaging Cherenkov (RICH) detectors.
The RICH photo detectors 
must be sensitive to single photons
with a  quantum efficiency $\int QE dE  \sim {\cal O}(1 eV)$ 
and provide spatial resolution with a granularity 
of about 2.5 x 2.5 mm$^2$ over  a large area of 
$\sim 3 \rm m^2$. 
The photo detectors must work in the magnetic 
fringe fields due to the LHCb dipole magnet and must cope with traversing charged particles.

\section{Multianode Photo Multipliers}
The multianode photo multiplier tube (MaPMT) consists of an array of
64 square anodes each with its own metal dynode chain  incorporated into a single
vacuum tube. 
The pixels have an area of $2.0
\times 2.0$ mm$^2$ and are separated by 0.3~mm gaps.
The MaPMT, manufactured by Hamamatsu,
has a 0.8~mm thick UV-glass window which transmits light
down to a wavelength of 200 nm.  
The photons are converted in a Bialkali photo cathode with a quantum efficiency
of maximum 22\% at 380 nm.  
The mean gain of the 12-stage dynode chain is about
$3 \times 10^5$ when operated at a voltage of 800 V.
\begin{figure}[htb]
\begin{center}
\mbox{\epsfig{file=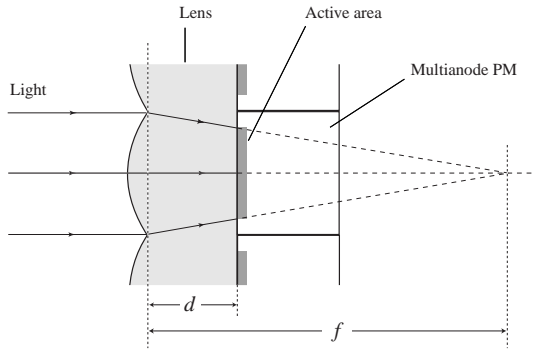,width=0.45\textwidth}}
\caption{Schematic view of lens/MaPMT system.
The focusing of normally incident light is illustrated. 
\label{fig:focus}}
\end{center}
\end{figure}

The ratio of the sensitive
photo cathode area to the total MaPMT area including the outer casing is only
$\sim$ 48\%.  This geometrical coverage  can be increased by placing a single lens with
one refracting  and one flat surface in front of each close-packed
tube (Fig.~\ref{fig:focus}).  
If the distance $d$ of the refracting  surface with radius-of-curvature $R$ to the
photo cathode is chosen to be equal to $R$ the  demagnification factor is
$\approx 2/3$.  Over the full aperture of the lens, light at normal
incidence with respect to the photodetector plane is focused onto the
photo cathode, thus restoring full geometrical acceptance.

\section{R\&D Results}

\begin{figure}[htb]
\begin{center}
\mbox{\epsfig{file=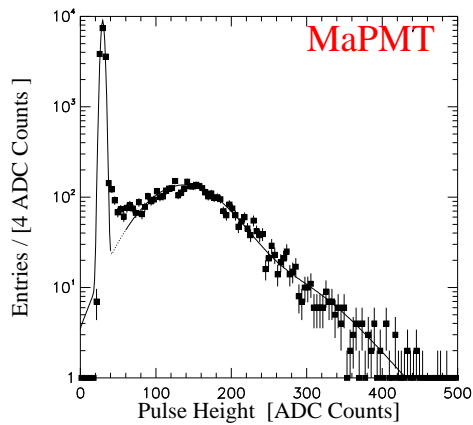,width=0.4\textwidth}}
\caption{Pulse height spectrum of a pixel.
\label{fig:amp_spectrum}}
\end{center}
\end{figure}

The pulse height spectrum for the MaPMT is shown in
Fig.~\ref{fig:amp_spectrum}, measured with a LED light source.  The
pedestal peak and the broad signal containing mostly one photo-electron
are clearly visible. The signal to pedestal width
ratio is 40:1.  
%
%
%

An array of 3x3 MaPMTs mounted onto 
the full-scale RICH\,1 prototype\cite{ref:vessel} 
has been tested in a beam at the CERN SPS facility.
The cathode voltage was set at $-1000$~V.
Quartz lenses were mounted onto the front face of each MaPMT.
The radiator was
gaseous CF$_4$ at a pressure of 700 mbar.
The data were recorded with a pipelined electronic
read-out system based on   
the APVm chip\cite{ref:apvm} and running at LHC speed (40 MHz).

The data analysis included 
a common-mode baseline subtraction on a event-by-event basis. 
With the pipelined read-out electronics cross-talk
was observed.  This has been investigated using LED runs  and
several sources -all in the electronics - were identified.  
The cross-talk is
removed by rejecting signals in a pixel if there is a larger signal in one
of its cross-talk partner pixels.  Genuine double hits are lost by this
procedure and the photon yield is corrected for it.

\begin{figure}[t]
\begin{center}
\mbox{\epsfig{file=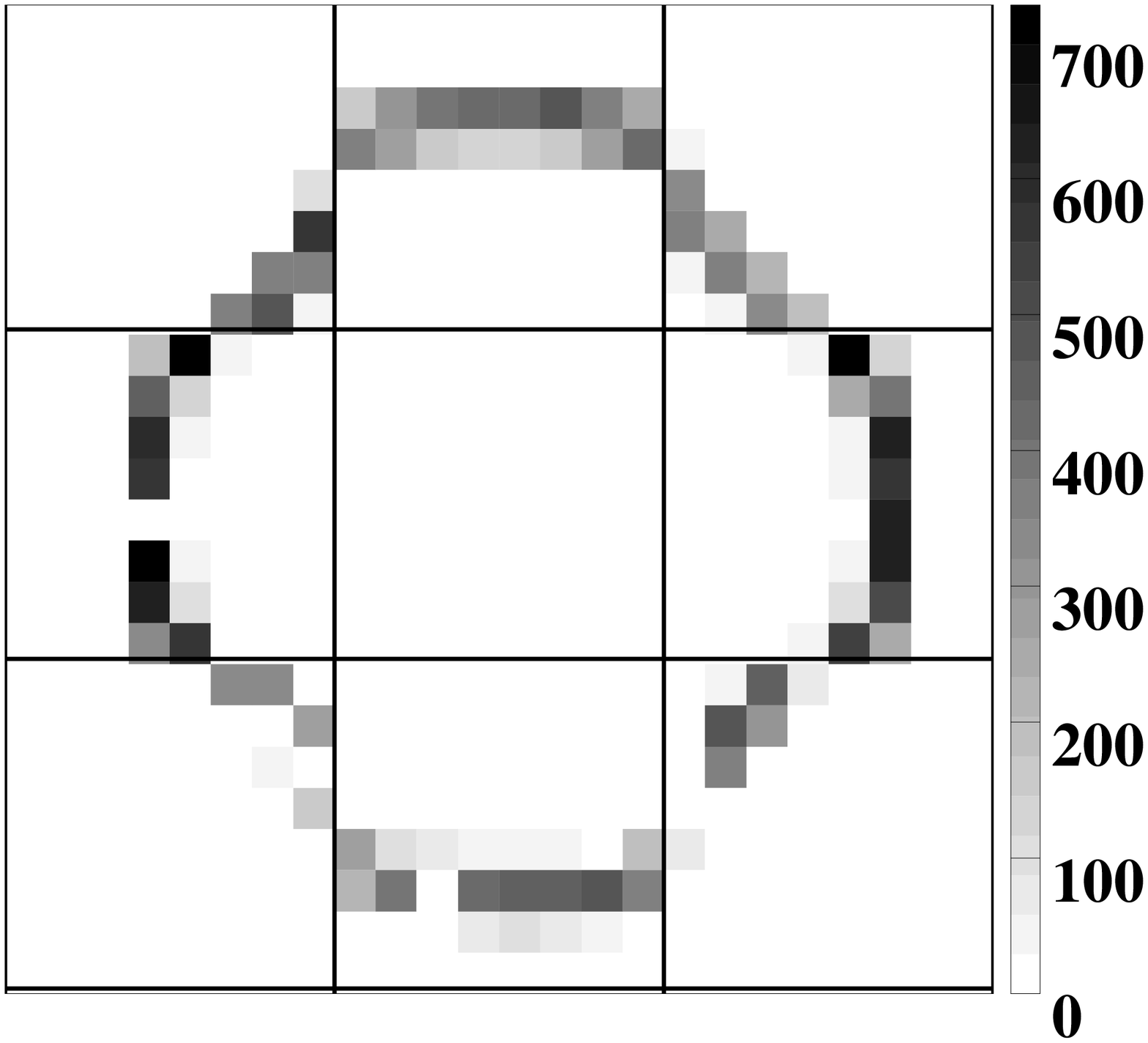,width=0.24\textwidth}
  \epsfig{file=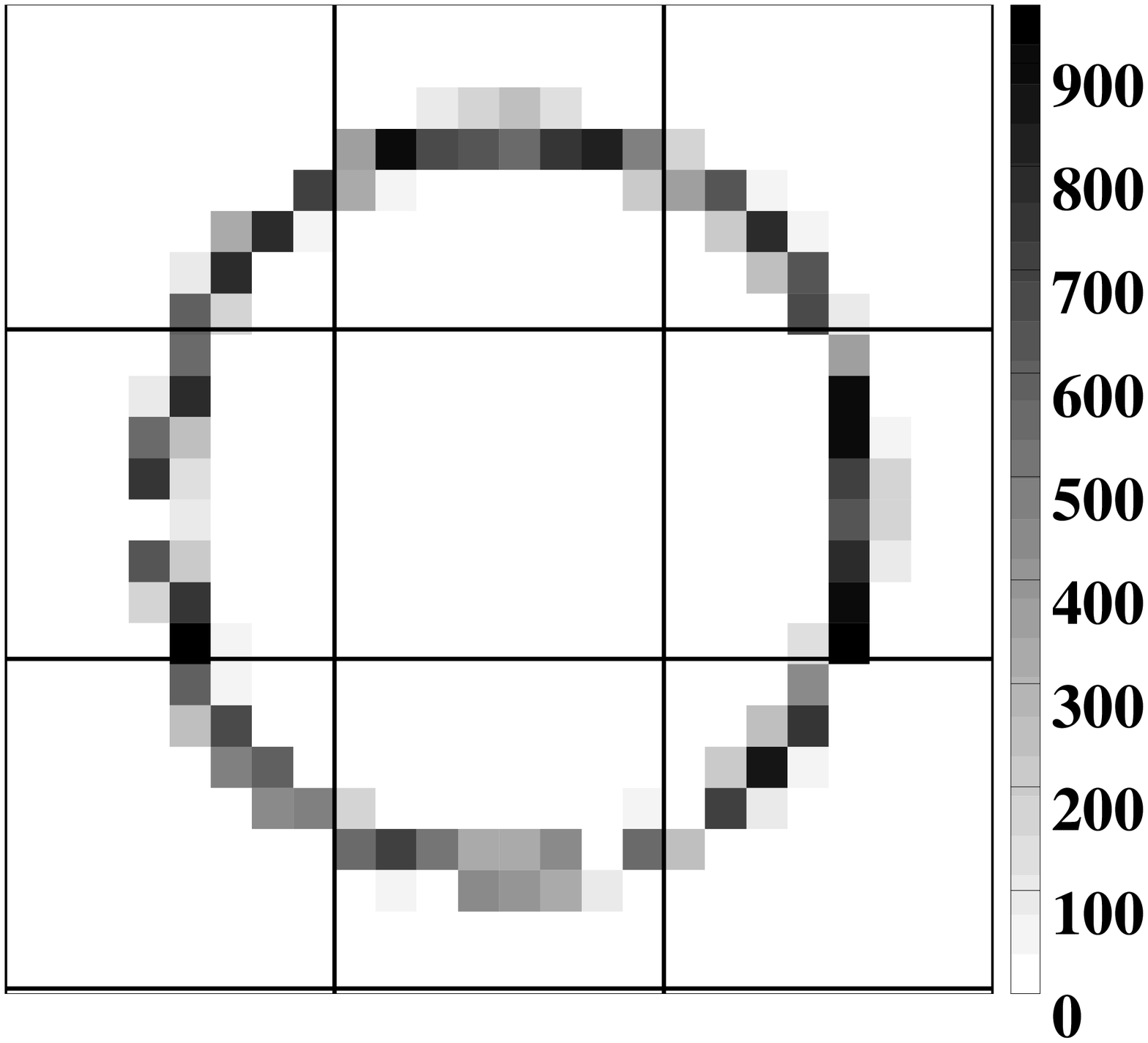,width=0.24\textwidth}}
\caption{Cherenkov ring measured with the 
MaPMT array with (right plot) 
and without (left plot) quartz lenses mounted in front of the tubes.
\label{fig:run2634}}
\end{center}
\end{figure}
The integrated signals of
two runs of 6000 events each are shown in Fig.~\ref{fig:run2634}, one
with and one without the lenses in front of the MaPMTs. The Cherenkov ring
is clearly visible and the effect of the lenses is nicely demonstrated.
The gain in in photo electrons by employing the lenses is 45\%. 
The background is small.
We measure $6.51 \pm 0.34$  photo electrons which is
is in good agreement with a full Monte Carlo  simulation.

We have also exposed the MaPMT/lens to charged particles. The measured response
is used to model the background in LHCb.
The sensitivity of the MaPMT to magnetic fields has been studied. The MaPMT 
is affected by longitudinal magnetic fields but can be 
effectively shielded from the expected field strengths with a $\mu$-metal structure. 

\section{Conclusions}

We have successfully tested a 3x3 array of MaPMTs.
Cherenkov light can be detected over the full area
of closely packed tubes
by means of quartz lenses focusing the light
onto the sensitive area of the device. 
We have demonstrated  that the MaPMT meets the performance requirements 
for charged particle identification in the LHCb experiment. 
The MaPMT has been chosen as the backup photo detector for LHCb.

\section*{Acknowledgments}
I thank all my MaPMT collaborators for their excellent work
presented here.

\end{document}